\documentclass{iau}
\usepackage{graphicx}

\title[Accretion, disks, and magnetic activity in the TW\,Hya association] %% give here short title %%
{Accretion, disks, and magnetic activity in the TW\,Hya association}

\author[B.Stelzer, A.Frasca \& J.M. Alcal\`a]   %% give here short author list %%
{B. Stelzer$^1, $
 A. Frasca$^2$
 \and J.M. Alcal\`a$^3$}

\affiliation{$^1$ INAF - Osservatorio Astronomico di Palermo, Piazza del Parlamento 1,
90134 Palermo, Italy \\ 
email: {\tt stelzer@astropa.inaf.it} \\[\affilskip]
$^2$ INAF - Osservatorio Astrofisico di Catania, Via Santa Sofia 78, 95123 Catania, Italy \\
$^3$ INAF - Osservatorio Astronomico di Capodimonte, Via Moiariello, 16, 80131 Napoli, Italy \\
}

\pubyear{2015}
\volume{314}  %% insert here IAU Symposium No.
\pagerange{xxx--xxx}
\jname{Young stars \& Planets Near the Sun}
\editors{J.H. Kastner, B.Stelzer \* S.A. Metchev, eds.}
\begin{document}

\maketitle

\begin{abstract}
We present new photometric and spectroscopic data for the M-type members of the TW\,Hya
association 
%reaching down into the brown dwarf regime 
with the aim of a comprehensive study of 
accretion, disks and magnetic activity at the critical age of $\sim 10$\,Myr 
where circumstellar matter disappears. 

\end{abstract}

\firstsection % if your document starts with a section,
\section{Introduction}\label{sect:intro}

Accretion, outflows and magnetic activity are key features driving young
stellar evolution. All these phenomena are linked in a complex feedback mechanism
where the activity from the stellar chromosphere and corona 
influences the evolution of disks and planet formation while mass accretion
from the disk to the star in turn may act as a heating agent on the structure 
of the outer stellar atmosphere. The details of this mechanism, how it depends on
stellar mass, and how it evolves with time is, thus, of utmost importance for an 
understanding of pre-main sequence evolution. 

At a distance of $\sim 50$\,pc the TW Hya association (TWA) represents one of the
most easily accessible laboratories for low-mass star formation. Moreover, 
its age of $\sim 8$\,Myr places the TWA at a crucial evolutionary phase where
disks dissipate and the accretion/outflow process comes to a halt. 

We are performing a broad-band (350-2500nm) mid-resolution spectroscopic
survey with X-Shooter@VLT. These observations give access to a rich database of emission 
lines which probe accretion, outflows and magnetic activity. 
%The data is complemented by archival X-ray and UV data of the same stars
%and compared to studies of M stars from nearby star forming regions and the 
%field which represent younger (1 Myr) and older (Gyrs) evolutionary phases, respectively. 
%Moreover, the X-Shooter spectroscopy enables a detailed characterization of the young 
%(sub)stellar objects. This includes an accurate assessment of their fundamental parameters, 
%kinematics, and rotation, parameters which are important for the interpretation of
%the activity/accretion/outflows properties.
In addition, we have carried out a optical/near-infrared (NIR) monitoring 
campaign of both stars in the wide binary system TWA\,30 to study photometric variability
related to their circumstellar environments. 
%The two objects are of similar spectral type (M4/5), 
%both have nearly edge-on disks, but show very different photometric variability attributed
%to different origins related to their circumstellar disks. 

\section{Chromospheric properties of TWA stars}\label{sect:activity}

\begin{figure}
% \vspace*{-2.0 cm}
\begin{center}
\parbox{14cm}{
\parbox{7cm}{
% FIGURE COPIED FROM: /Users/bstelzer/conf/FIGURES/
 \includegraphics[width=7cm]{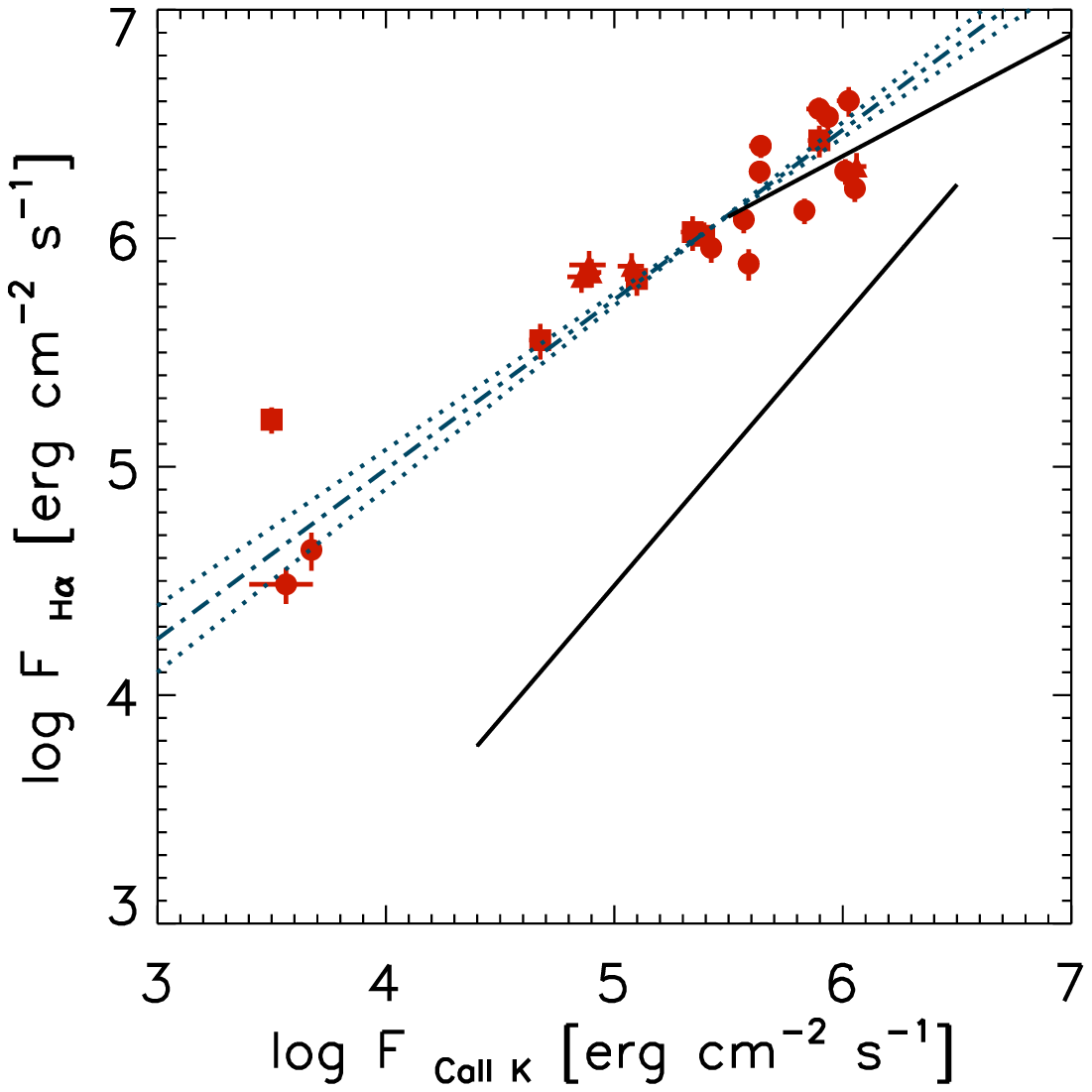}
 }
 \parbox{6.5cm}{
   \includegraphics[width=6.0cm]{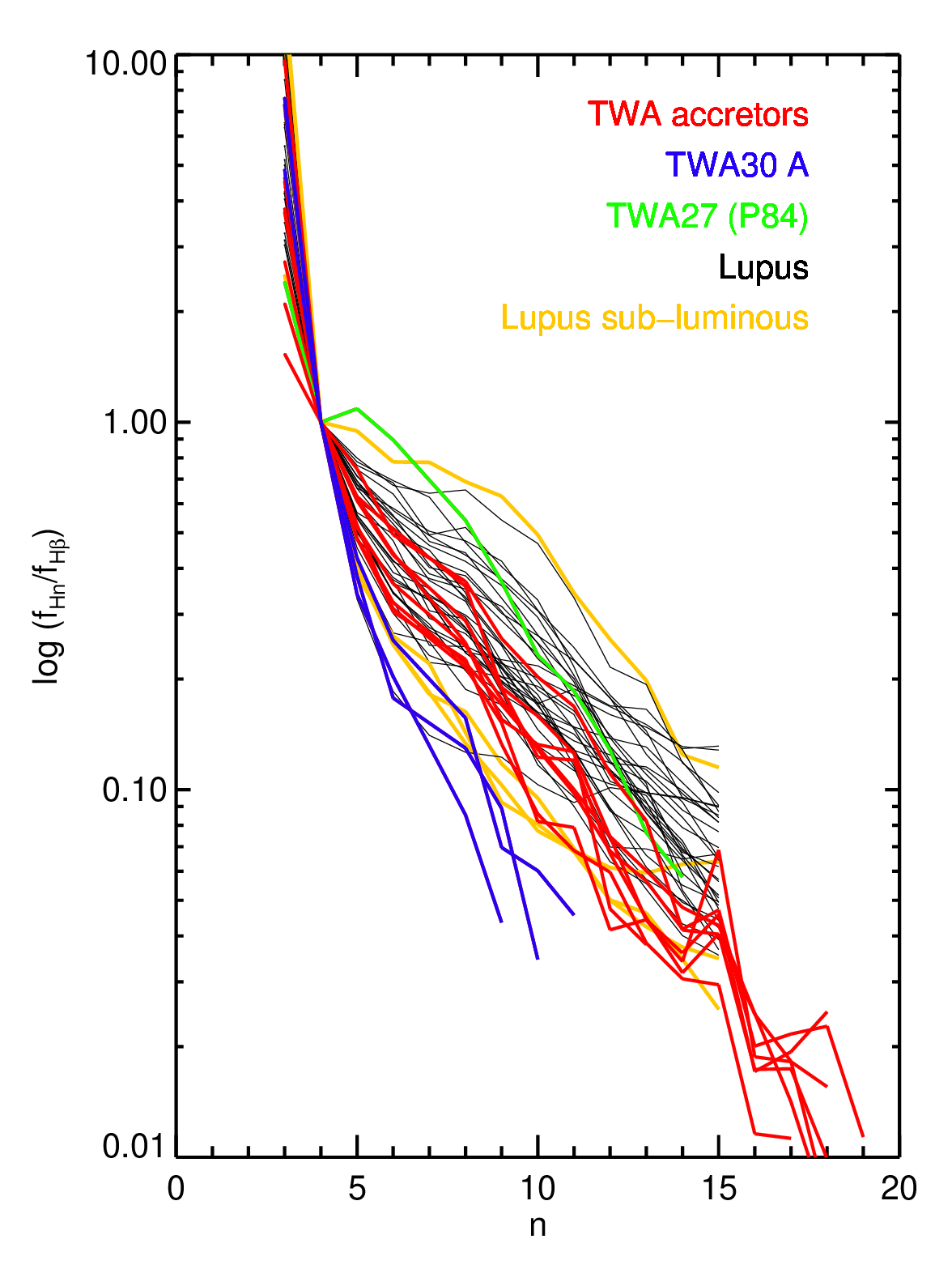}
 }
 }
% \vspace*{-1.0 cm}
 \caption{
 {\it (Left)} Flux-flux relation between chromospheric 
 H$\alpha$ and Ca\,II\,K emission for Class\,III stars; %pre-main sequence stars; 
 upper/lower black lines denote the ``active''/``inactive'' branches 
 defined by the main-sequence sample of \cite{Martinez-Arnaiz10} and the blue line is 
 the best-fit to the pre-main sequence sample. 
 {\it (Right)} Balmer decrements for accretor candidates in TWA (colored curves) compared
 to those for accretors in Lupus (black curves; Antoniucci et al. 2015, in prep). 
 %(right) Flux ratio as function of effective temperature.
 } 
\label{fig:fluxflux}
\end{center}
\end{figure}

%Traditionally, the chromospheres of late-type stars are studied through their strongest
%emission lines, H$\alpha$ and Ca\,II\,HK emission. Our knowledge on the whole
%emission line spectrum is more elusive as a result of the limited spectral range and 
%sensitivity of most available spectrographs.
%We intend to reduce this gap with a comprehensive spectroscopic study of the chromospheric
%emission line spectrum of a sample of non-accreting pre-main sequence stars (Class\,III sources). 
We have analyzed X-Shooter/VLT spectra of $24$ non-accreting pre-main sequence stars,
so-called Class\,III sources, from three nearby 
star-forming regions ($\sigma$\,Orionis, Lupus\,III, and the TWA). 
%We determined the effective temperature, surface gravity, rotational 
%velocity, and radial velocity by comparing the observed spectra with synthetic BT-Settl model 
%spectra. We investigated
%in detail the emission lines emerging from the stellar chromospheres and combined these
%data with archival X-ray data to allow for a comparison between chromospheric and coronal 
%emissions.
Of particular interest here are flux-flux relations between 
activity diagnostics that probe different atmospheric layers (from the lower chromosphere
to the corona). We examine such relations for the first time for pre-main sequence stars. 
For main-sequence late-type stars the fluxes
of chromospheric emission lines follow power-law relations, with the exception of a subgroup
of M-type stars which define a separate ``active'' branch  
(\cite[Martinez-Arn\'aiz et al. 2010]{Martinez-Arnaiz10}).
Fig.~\ref{fig:fluxflux} (left) shows that all the Class\,III stars from our sample lie on this  
``active'' branch, and -- spanning a range of masses throughout the full M spectral 
sequence -- they extend it to lower fluxes. We conclude that chromospheric
flux-flux relations can serve as a tool for confirming the youth of low-mass stars.  
More detailed investigations covering ages between the $\sim 10$\,Myr of TWA and the 
$\geq 1$\,Gyr of main-sequence field stars are required to explore this new age diagnostic. 

Flux ratios between individual emission lines of the Class\,III sources 
show a smooth dependence on the effective temperature.  
%(e.g. Fig.~\ref{fig:fluxflux} right). 
\cite{Meunier09} found that in the solar
chromosphere the different behavior of H$\alpha$ and Ca\,II\,H\&K is due to their link to 
different structures, so-called filaments and plages. 
However, the observed spread in Fig.~\ref{fig:fluxflux} and in the H$\alpha$/Ca\,II\,K
flux ratio for given $T_{\rm eff}$ suggests that a third 
parameter, in addition to age and %spectral type (or 
$T_{\rm eff}$, 
determines the chromospheric structure of late-type stars. 
%The Balmer decrements can roughly be reproduced by an NLTE radiative transfer model devised 
%for another young star of similar age. Future, more complete chromospheric model grids can
%be tested against this data set. 

H$\alpha$ and Ca\,II\,H\&K are the best-studied emission lines
in chromospheres of late-type stars. Our knowledge on the whole
emission line spectrum is more elusive as a result of limited spectral range and 
sensitivity of most available spectrographs. This gap can be closed with wide-band
spectroscopic studies, such as our 
comprehensive X-Shooter study of Class\,III sources 
(see \cite[Stelzer et al. 2013]{Stelzer13a}).  

\section{Accretion measurements for TWA members}\label{sect:accretion}

We have obtained X-Shooter spectra for $15$ ``potentially accreting'' TWA members. 
Here we concentrate on measurements of mass accretion rates ($\dot{M}_{\rm acc}$).
We consider as ``potential accretors'' all TWA stars previously known to be accreting 
and further TWA members with evidence of IR (disk) excess in published
photometry or a Balmer jump in the X-Shooter spectra. 
This sample excludes the Class\,III stars discussed in Sect.~\ref{sect:activity}. 

Two methods are used to measure mass accretion rates from the X-Shooter spectra.
First, we determine the Balmer jump -- if any -- by modelling the UVB spectrum with
a combination of a non-accreting Class\,III template and a hydrogen slab model representing
emission from the accretion column as described by \cite{Manara13}. This yields
the accretion luminosity, $L_{\rm acc}$. 
Secondly, we use the empirical calibrations between fluxes of individual emission lines
and the $L_{\rm acc}$ values measured with the slab modelling derived by \cite{Alcala14}
to obtain $\dot{M}_{\rm acc}$ from observed line fluxes. 

We compare our preliminary results for $\dot{M}_{\rm acc}$ as a function of stellar mass ($M_*$)
to the analogous study for the Lupus star forming region by \cite{Alcala14}, and find that
the TWA stars display smaller $\dot{M}_{\rm acc}$ at given mass. However, the slope in 
the $M_* - \dot{M}_{\rm acc}$ diagram is the same as for Lupus. This indicates that the 
TWA disks represent an evolved version of the same type of disks that are present in Lupus.

In Fig.~\ref{fig:fluxflux} (right) we compare the Balmer decrements of the TWA accretor
sample to those of the accretors in Lupus (see Antoniucci et al. 2015, in prep.).
The TWA stars tend to have slightly steeper decrements than the younger
stars in Lupus. The physical conditions determining the shape of the decrements 
will be investigated by comparison to decrements predicted from the 
radiative transfer models by \cite{Kwan11}.

\section{Multiband photometric monitoring of the TWA\,30 binary}\label{sect:disks}

TWA\,30A and 30B are nearly equal-mass components of a wide binary 
(spectral types M5 and M4; $80^{\prime\prime}$ separation) 
discovered and confirmed as kinematic TWA members 
by \cite{Looper10a} and \cite{Looper10b}. They 
are among the nearest ($\sim 50$\,pc) stars with 
signatures of disks, %(infrared excess emission), 
mass accretion and outflow activity. % (permitted and forbidden emission lines (FELs)). 
The disks of both stars are likely seen (nearly) edge-on: 
(i) The forbidden emission lines (FELs)   
have only small radial velocity shifts indicating that their jets are oriented in the plane
of the sky; (ii) the FELs are 
much stronger than the accretion signatures suggesting that
the accretion columns are partially hidden by parts of the circumstellar disk. 
%(The outflows are 
%formed at some distance from the star and remain unobscured.)
%The optical/near-infrared (NIR) spectra of both stars can best be 
%understood if their circumstellar disks are seen edge-on. 
%These observations imply the presence of circumstellar disks seen (nearly) edge-on.

\begin{figure} %[b]
% \vspace*{-2.0 cm}
\begin{center}
\parbox{14cm}{
\parbox{7cm}{
% FIGURES COPIED FROM: /Users/bstelzer/proposals/TNG/AOT32/SUBM/
 \includegraphics[width=7cm]{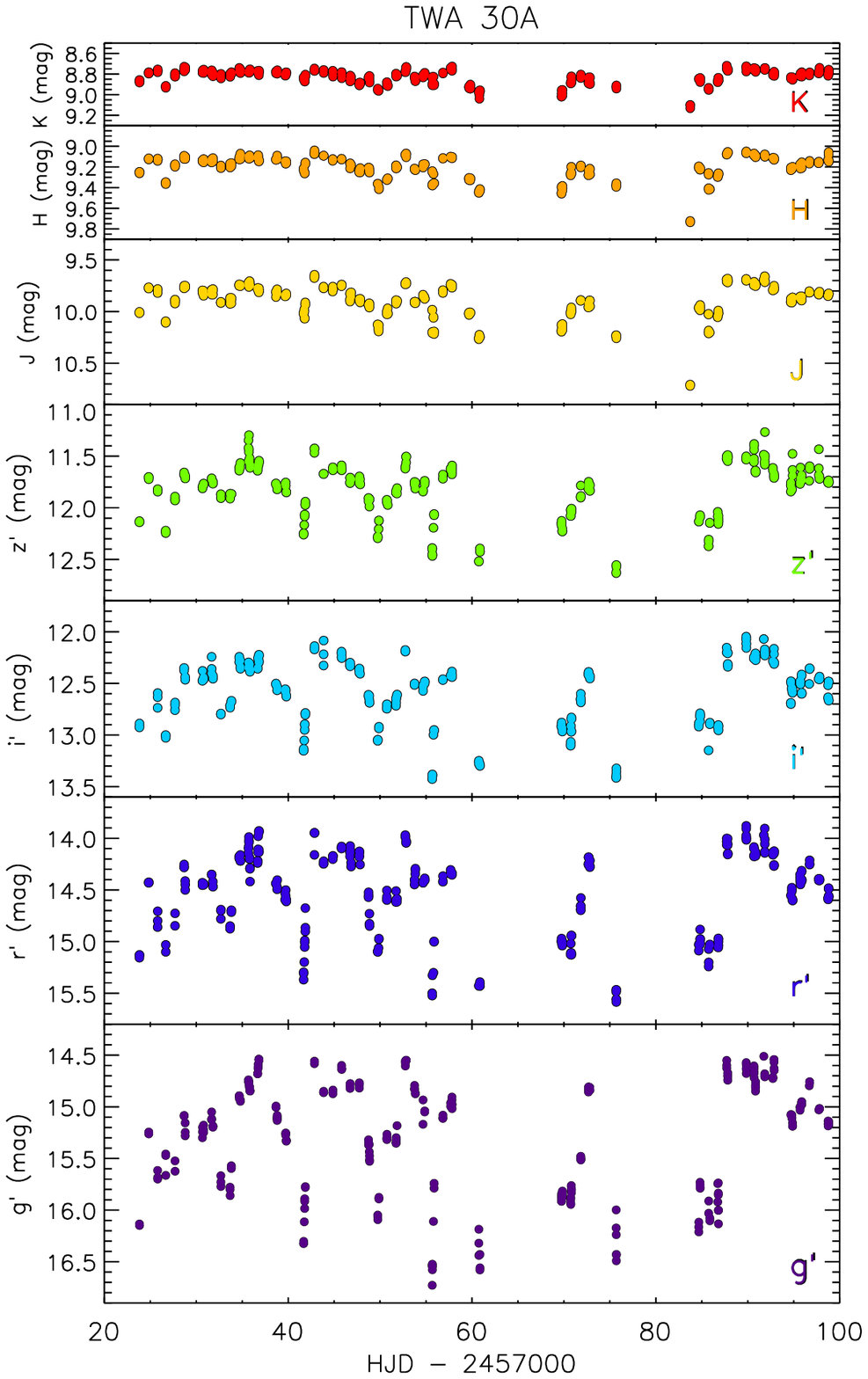} 
 }
 \parbox{7cm}{
 \includegraphics[width=7cm]{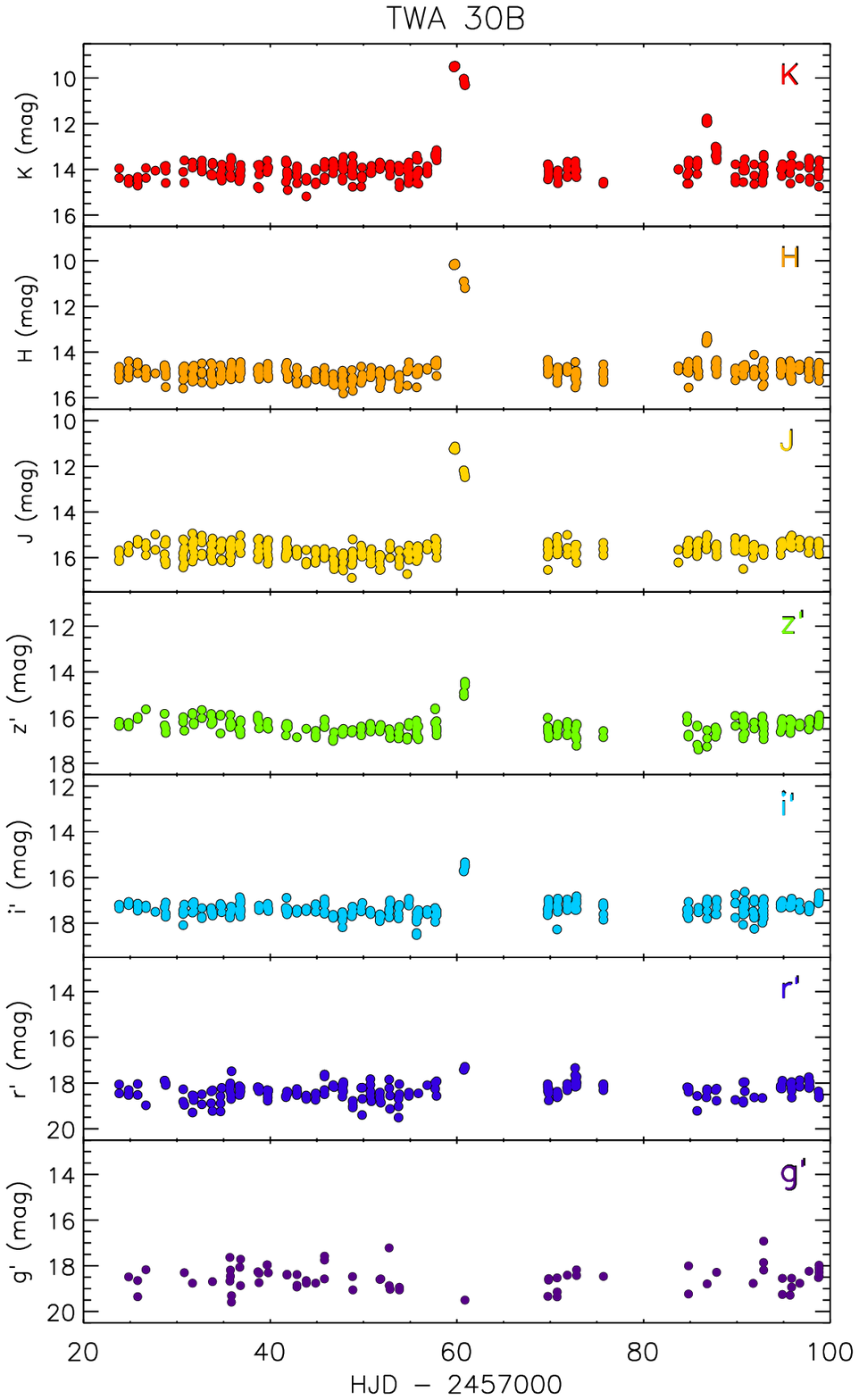} 
 }
 }
% \vspace*{-1.0 cm}
 \caption{REM multi-band optical/NIR lightcurves of TWA\,30A (left) and TWA\,30B (right).
}
\label{fig:twa30_REMlcs}
\end{center}
\end{figure}

In the discovery papers, \cite{Looper10a} and \cite{Looper10b} showed with 
multi-epoch low-cadence spectroscopy that both stars have strongly variable NIR colors. 
%No simultaneous optical and NIR data was available to bolster these interpretations. 
To constrain the time-scales of the variability over its full optical/NIR spectral energy
distribution we have carried out a photometric monitoring
campaign with the robotic $60$\,cm telescope REM on La Silla operated/owned by the Italian 
Istituto Nazionale di Astrofisica (INAF).  
The REM is equipped with a NIR camera, REMIR, and an optical camera, ROSS\,2, fed by the 
same telescope and providing simultaneous photometry in seven filters, g'r'i'z'JHK'. 
The REM lightcurves of TWA\,30A and 30B are shown in Fig.~\ref{fig:twa30_REMlcs}.

TWA\,30A shows quasi-periodic dimmings throughout the monitoring. 
The lightcurves observed in the seven filters are highly
correlated and the depth increases systematically for bluer bands. The NIR colors 
show that the data points are roughly aligned along the reddening vector. 
However, during the whole 3-month monitoring campaign the extinction did not reach the high 
($\sim 10$\,mag) values inferred from previous spectroscopy. 
%The changes seen in the depth of the dips indicates that the
%amount of optical depth of the occulting material is highly variable. 
Similar lightcurves have been observed from $9$ stars in the NGC\,2264 star forming region 
based on a long CoRoT and Spitzer monitoring campaign (Stauffer et al. 2015). 
The occurrence of such periodic dimmings can be interpreted as dust structures 
levitated above the
disk and temporarily occulting the star light as they pass in front of the line of sight.

TWA\,30B was for most of the time during our REM monitoring observed in a faint state with 
occasional ``bursts'' of several magnitudes. 
Contrary to TWA\,30A, the brightness variations of TWA\,30B 
are much stronger in the NIR than at optical bands. The ``bursts'' observed in the REM
lightcurve are likely responsible for the NIR excess seen in the previous spectroscopic
data, which was modelled by Looper et al. (2010b) as cool ($\sim 700$\,K) blackbody
emission. A physical interpretation
for this excess NIR emission could be starlight reprocessed from a small portion of the disk
that rotates in and out of the line-of-sight.
For TWA\,30B the NIR photometry has a much flatter slope in the $J-H$ vs $H-K_s$ diagram
than TWA\,30A, and NIR colors measured for the ``bursts'' seen during our 
REM monitoring are different from those inferred from previous spectrophotometry
by \cite{Looper10b}. 
%The star is too faint to constrain the "quiescent" emission to sufficient
%precision for calculating colors during its faint state. 

\begin{discussion}

\discuss{Rodriguez}{Did you find periodicities in the REM lightcurves of TWA\,30 ? }

\discuss{Stelzer}{Yes, indeed. The dips of TWA\,30A are quasi-periodic
with $P \sim 8$\,d.  Assuming corotation, the observed period implies that the dust is 
located at a distance of $\sim 0.2$\,AU from the star (i.e. in the inner disk) and that 
its azimuthal position is roughly consistent over time. The variable depth of the dips
indicates that the amount or optical depth of the occulting material is changing in time. 
}

\end{discussion}

\end{document}